\documentstyle[11pt,axodraw]{article}

\begin{document}

\begin{center}
{\LARGE \bf General massive one-loop off-shell three-point functions}

\vspace{1.5cm}

\bigskip {\large A. T. Suzuki\footnote{E-mail:suzuki@ift.unesp.br}}{\large ,
E. S. Santos\footnote{E-mail:esdras@ift.unesp.br}}

Instituto de F\'{\i}sica Te\'{o}rica

Universidade Estadual Paulista

Rua Pamplona 145

S\~{a}o Paulo SP - 01405-900

Brazil

\vspace{0.5cm}
{\large A. G. M. Schmidt\footnote{E-mail:schmidt@fisica.ufpr.br}}

Universidade Federal do Paran\'{a}

Departamento de F\'{\i}sica

Caixa Postal 19044

Curitiba PR - 81531-990

Brazil

\vspace{1cm}

\begin{minipage}{14.5cm}
\centerline{\bf Abstract}

{In this work we compute the most general massive one-loop off-shell
three-point vertex in $D$-dimensions, where the masses, external
momenta, and exponents of propagators are arbitrary. This follows our
previous paper in which we have calculated several new hypergeometric
series representations for massless and massive (with equal masses)
scalar one-loop three-point functions, in the negative dimensional
approach.}
\end{minipage}
\end{center}

\bigskip Keywords: negative dimensional integration, one-loop triangle
diagram.

02.90+p, 12.38.Bx

\bigskip

\newpage

\section{Introduction}

In perturbative quantum field theories, obtaining analytic results for
the calculation of Feynman loop integrals with massless or massive
internal particles is often a hard task. Even more so if one is
considering external particles off the mass-shell
condition. Undoubtedly, in spite of the technical complexities
attached to such calculations, great advances in this field have been
accomplished \cite{anastasiou}-\cite{binoth}, and the inherent
difficulties overcome to the extent that now several two-loop and even
higher order Feynman integrals are already known. In some cases one
can generalize simple results to the case where one has an arbitrary
number of loops \cite{ach,ussyukina,belokurov}.

In the two-loop arena, Glover and co-workers studied several integrals
pertaining to $2\rightarrow 2$ scattering \cite{anastasiou,glover} and
references therein. They completed the whole task. Bern, Dixon and
co-workers also tackled such integrals in order to study other processes
such as $2\rightarrow 3$, $2\rightarrow 4$ and light by light scattering
\cite{bern-penta,bern-6pernas,freitas}. Recently, Smirnov, in the
three-loop arena presented the result for a scalar massless triple-box
using Mellin-Barnes approach \cite{smirnov-3loops}.

Thus, the present situation calls for the search and application of
sophisticated techniques to make manageable diagrams with growing
complexities leading to multi-loop integrals with a hefty load of
algebraic manipulations and mathematical tools. In this scenario, we
can witness considerable advances taking place in the calculation of
those multi-loop integrals. On the other hand, some of the ``simpler''
diagrams and associated loop integrals have not yet been fully
studied; the case of the triangle diagram being one such example. This
was the subject of our previous paper where we considered a one-loop
triangle integral with one, two and three massive particles in the
intermediate states (with equal masses), and arbitrary exponents of
propagators \cite{nossopaper}. We did show that there are dozens of
generalized hypergeometric series that represent such integrals,
but just a few of them are known in the literature \cite{davydy}.

Now we generalize those results in order to calculate scalar
three-point integrals with arbitrary masses and external momenta.  Our
results are given in terms of generalized hypergeometric functions,
most of them not known in the literature.


The outline for our paper is as follows: in section 2 we solve the
most general triangle integral and present several particular cases of
interest. The third section is devoted to our conclusions, and in the
appendices we put all the tables of results, and the definitions of
hypergeometric series we used throughout our paper.

\section{One-loop three-point function}
$ $

\begin{figure}[tbp]
\begin{center}
\vspace{25mm}
\begin{picture}(400,120)(0,-70)
\ArrowLine(50,0)(0,0)
\Line(80,52)(50,0)
\ArrowLine(80,52)(80,102)
\Line(110,0)(80,52)
\Line(110,0)(50,0)
\ArrowLine(160,0)(110,0)
\Text(25,10)[c]{$p$}
\Text(80,10)[c]{$k$}
\Text(135,10)[c]{$q$}
\Text(45,26)[c]{$p-k$}
\Text(100,77)[c]{$q-p$}
\Text(115,26)[c]{$q-k$}
\Text(80,-20)[c]{Fig.$1$}
\ArrowLine(260,40)(200,40)
\ArrowLine(380,40)(320,40)
\CArc(290,40)(30,0,360)
\Text(230,50)[c]{$p$}
\Text(290,80)[c]{$k$}
\Text(350,50)[c]{$p$}
\Text(290,0)[c]{$p-k$}
\Text(290,-20)[c]{Fig-$2$}
\end{picture}
\end{center}
\end{figure}

In this section, we calculate using NDIM the scalar integral of the
one-loop triangle diagram (see Fig. 1) with two independent external
momenta. Particular cases such as the massless case, one, two, and
three equal masses cases are not considered here because these are
treated in our previous work \cite{nossopaper}. Rather here we
consider other limiting cases such as when one has a vanishing
exponent of a given propagator. This said, we have that the
three-point function is given by the following scalar integral:
\begin{eqnarray}
J^{(3)}&=&J^{(3)}(a,b,c,D,p,q,m_{1},m_{2},m_{3})\nonumber \\
&=&\int \frac{d^{D}k}{
[k^{2}-m_{1}^{2}]^{a}[(k-p)^{2}-m_{2}^{2}]^{b}[(k-q)^{2}-m_{3}^{2}]^{c}}.
\label{a1}
\end{eqnarray}
 To implement the NDIM method we start off by considering the
 corresponding Gaussian integral
\begin{eqnarray}
I &=&I(p,q,m_{1},m_{2},m_{3})  \nonumber \\
&=&\int d^{D}k\exp \{-\alpha (k^{2}-m_{1}^{2})-\beta \lbrack
(k-p)^{2}-m_{2}^{2}]-\gamma \lbrack (k-q)^{2}-m_{3}^{2}]\},  \nonumber \\
\label{a2}
\end{eqnarray}
which can be rewritten in the form
\begin{eqnarray}
I&=&{\sum_{a,b,c=0}^\infty}(-1)^{a+b+c}\frac{\alpha ^{a}\beta ^{b}\gamma ^{c}}{
a!b!c!} \nonumber \\ 
&\times& \int d^{D}k\;
[k^{2}-m_{1}^{2}]^{a}[(k-p)^{2}-m_{2}^{2}]^{b}[(k-q)^{2}-m_{3}^{2}]^{c}\\  
&=& \sum_{a,b,c=0}^{\infty}(-1)^{a+b+c}\frac{\alpha^a\beta^b\gamma^c}{a!b!c!}\nonumber \\ 
&\times& J^{(3)}(-a,-b,-c,D,p,q,m_{1},m_{2},m_{3}).  \label{a3}
\end{eqnarray}
Note that after carrying out the analytic continuation to negative
values of $a,b$ and $c$ in (\ref{a3}) we have that
$J^{(3)}(-a,-b,-c,p,q,m_{1},m_{2},m_{3})$ is just the sought after
$D$-dimensional massive Feynman integral for the three-point function in (\ref{a1}).


Performing the $D$-dimensional Gaussian integral (\ref{a2}), we have
\begin{eqnarray}
J &=&\pi ^{D/2}(-1)^{-a-b-c}a!b!c!{\sum_{j_{1},...,j_{9}=0}^\infty }
\frac{\Gamma (1-j_{1}-j_{2}-j_{3}-D/2)}{
j_{7}!j_{8}!j_{9}!}  \nonumber \\
&&\times \frac{(-p^{2})^{j_{1}}}{j_{1}!}\frac{(-q^{2})^{j_{2}}}{j_{2}!}\frac{
(-r^{2})^{j_{3}}}{j_{3}!}\frac{(m_{1}^{2})^{j_{4}}}{j_{4}!}\frac{
(m_{2}^{2})^{j_{5}}}{j_{5}!}\frac{(m_{3}^{2})^{j_{6}}}{j_{6}!}  \label{a5}
\end{eqnarray}
with the following constraint equations
\begin{eqnarray}
a &=&j_{1}+j_{2}+j_{4}+j_{7}  \label{a6} \\
b &=&j_{1}+j_{3}+j_{5}+j_{8}  \label{a7} \\
c &=&j_{2}+j_{3}+j_{6}+j_{9}  \label{a8} \\
\frac{D}{2} &=&-j_{1}-j_{2}-j_{3}-j_{7}-j_{8}-j_{9}  \label{a9}
\end{eqnarray}
where the constraints (\ref{a6}-\ref{a8}) come from comparing the
powers of $\alpha ,\beta $ and $\gamma$ with the sum in the indices
$j_{7},j_{8}$ and $j_{9}$. The last one, (\ref{a9}), is related to the
polynomial expansion $(\alpha +\beta+\gamma )^{-D/2-j_1-j_2-j_3}$.

Now, since a system of four linear algebraic equations with
nine variables can only be solved if we leave five free indices, the
final result will be given in terms of a five-fold series. These
remaining sums can be constructed in one hundred and twenty six
different ways, each one with five free indices chosen. This number
comes out quickly from the combinatorics
\begin{equation}
C_{9}^{5}=\frac{9!}{5!4!}=126.  \label{a10}
\end{equation}
Among these 126 solutions, 45 have trivial solutions (the determinant
of the system is zero) while 12 vanishes due to Pochhammer's factors
in the coefficients of the series of the type
\begin{equation}
(0)_{a+D/2} = \frac{\Gamma (a+D/2)}{\Gamma (0)}=0\;, 
\end{equation} 
so that only 69 are non-trivial and independent. They are given in
terms of four hypergeometric and hypergeometric-type series of the
powers of the external momenta and internal masses (see Appendices A
and C).

In order to make clearer our notations, definitions and usage of
symbols, we pick up a specific solution to show how the result is
written down in the corresponding tables in the Appendices. From
Tables 1 and 2 of Appendix C we read off the solution number 63 as
follows:

\begin{eqnarray}
J^{(3)}&=&J^{3}(a,b,c,p,q,r,m_1,m_2,m_3,D)\nonumber \\
&=&\pi^{D/2}(-m_{3}^{2})^{a+b+c+D/2}(D/2)_{a+b}(-c)_{-a-b-D/2} \nonumber \\
&\times& \Psi _{3}\left[ \left.
\begin{array}{l}
-\sigma,\;-a,\;-b \\
1-\sigma+c,\;D/2
\end{array}
\right |-\frac{p^{2}
}{m_{3}^{2}};\;\frac{q^{2}}{m_{3}^{2}};\;\frac{r^{2}}{m_{3}^{2}};\;\frac{m_{1}^{2}
}{m_{3}^{2}};\;\frac{m_{2}^{2}}{m_{3}^{2}}\right],
\end{eqnarray}
where the hypergeometric function $\Psi_3$ is defined in 
Appendix A and is valid in the kinematic region where 
$
m_3 \neq  0
$
and 
$
p^2 <  m_3^2\,,\;\;q^2  <  m_3^2\,,\;\;r^2  =  (q-p)^2 <  m_3^2\,,\;\;m_1^2  <  m_3^2\,,\;\;m_2^2  <  m_3^2\;.
$

Coefficients of solutions contain the Pochhammer
symbol defined by 
\begin{equation}
(x)_y=\frac{\Gamma(x+y)}{\Gamma(x)}\;.
\end{equation}

\subsection{One-loop two-point function}
$ $

Here we present the two-point function obtained from Feynman
diagram depicted in (see Fig.2) as a particular case of the three-point
function. To achieve this we take $c=0$ in (\ref{a1}), so that 

\begin{eqnarray}
J^{(2)}&=&J^{(2)}(a,b,D,p,m_{1},m_{2})\nonumber \\
&=&\int \frac{d^{D}k}{
[k^{2}-m_{1}^{2}]^{a}[(k-p)^{2}-m_{2}^{2}]^{b}}.
\end{eqnarray}

Letting the $c$ exponent to be zero implies also that
$j_2=j_3=j_6=j_9=0$ in (\ref{a5}) and (\ref{a8}).  In this case, we
have now three constraint equations, namely
(\ref{a6},\ref{a7},\ref{a9}) and five variables. From this, the
combinatorial analysis gives us $C_{5}^{3}=10$ different
possibilities for the systems of linear algebraic equations. Three of
them have vanishing determinant, so that we are left over with seven
non trivial solutions. These are grouped into three sets, one with
three solutions and two with two solutions each, according to the
kinematical configuration of the variables defined by the external
momenta (see Appendix B). Performing the analytic continuation to
$D>0$ and $a,b<0$ we get the three sets of solutions for the relevant
Feynman integral given by
\begin{eqnarray}
\label{2pointa}
J^{(2)}&=&J_1+J_2+J_3,\\
J^{(2)}&=&J_4+J_5, \label{2pointb} \\
J^{(2)}&=&J_6+J_7,\label{2pointc} 
\end{eqnarray}
where $J_1, J_2,...,J_7$ are listed in Tables 1 and 2 of 
Appendix B. The last solution above, i.e., $J^{(2)}=J_6+J_7$ reads
\begin{eqnarray}
J^{(2)}&=&\pi^{D/2}(-m_1^2)^{a+D/2}(-m_2^2)^b(-a)_{-D/2}\nonumber \\
&&\times F_{4}\left[\left.
\begin{array}{l}
D/2,-b \\
D/2,1+a+D/2
\end{array}\right
|\frac{p^2}{m_2^2};\frac{m_1^2}{m_2^2}\right] \nonumber \\
&+&\pi^{D/2}(-m_2^2)^{a+b+D/2}(-b)_{-a-D/2}(D/2)_{a} \nonumber \\
&&\times F_{4}\left[\left.
\begin{array}{l}
-a,-a-b-D/2 \\
D/2,1-a-D/2
\end{array}\right
|\frac{p^2}{m_2^2};\frac{m_1^2}{m_2^2}\right]
\end{eqnarray}

All the sets of solutions in (\ref{2pointa})-(\ref{2pointc}) agree with the results previously known in the literature \cite{davydy}.

\section{Conclusion}
$ $

In this paper we have used the NDIM approach to evaluate the general
massive one-loop triangle diagram. Diagrams of this type are relevant
to the study of vertex corrections in QED and QCD, scalar
electrodynamics, electroweak interactions and so on. We have obtained
sixty nine solutions, each one pertaining to a specific distinct
kinematical region, that show us all the different ways in which the
three-point function may be expressed. These results are represented
by hypergeometric-type functions of five variables. Particular cases
of our results can be seen in \cite{nossopaper,davydy}.  The two-point
function obtained as a limiting case $c=0$ from our results are in
agreement with \cite{davydy}. Therefore, the NDIM have revealed itself
as a good alternative technique to the computation of Feynman
integrals.

\bigskip

\appendix

\section{Hypergeometric functions used}
$ $

The Appel hypergeometric function and the hypergeometric-type
functions of five variables, used in this paper, satisfy the
differential equations shown in \cite{nossopaper} and are listed
below.

\begin{eqnarray*}
F_{4}\left[\left.
\begin{array}{l}
x_{1},x_{2} \\
x_{3},x_{4}
\end{array}\right
|z_{1};z_{2}\right] &=&{\sum_{j_{1},j_{2}=0}^{\infty} }\frac{
(x_{1})_{j_{1}+j_{2}}(x_{2})_{j_{1}+j_{2}}}
{(x_{3})_{j_{1}}(x_{4})_{j_{2}}}\frac{z_{1}^{j_{1}}}{j_{1}!}\frac{z_{2}^{j_{2}}}{j_{2}!}, \\
\Psi _{1}\left[\left.
\begin{array}{l}
x_{1},x_{2},x_{3} \\
x_{4},x_{5}
\end{array}\right
|z_{1};z_{2};z_{3};z_{4};z_{5}\right] &=&{\sum_{j_{1},..,j_{5}=0}^{\infty} }\frac{
(x_{1})_{j_{1}+j_{2}+j_{3}}(x_{2})_{j_{1}+j_{2}-j_{4}}(x_{3})_{j_{1}+j_{3}-j_{5}}
}{(x_{4})_{j_{1}-j_{4}-j_{5}}(x_{5})_{j_{1}+j_{2}+j_{3}-j_{4}-j_{5}}} \\
&&\times \frac{z_{1}^{j_{1}}}{j_{1}!}\frac{z_{2}^{j_{2}}}{j_{2}!}\frac{
z_{3}^{j_{3}}}{j_{3}!}\frac{z_{4}^{j_{4}}}{j_{4}!}\frac{z_{5}^{j_{5}}}{j_{5}!
}, \\
\Psi _{2}\left[\left.
\begin{array}{l}
x_{1},x_{2},x_{3} \\
x_{4},x_{5}
\end{array}\right
|z_{1};z_{2};z_{3};z_{4};z_{5}\right] &=&{\sum_{j_{1},..,j_{5}=0}^{\infty} }\frac{
(x_{1})_{j_{1}+j_{2}+j_{5}}(x_{2})_{j_{3}+j_{4}+j_{5}}(x_{3})_{j_{1}+j_{2}-j_{4}}%
}{(x_{4})_{j_{2}-j_{3}-j_{4}}(x_{5})_{j_{1}+j_{3}+j_{5}}} \\
&&\times \frac{z_{1}^{j_{1}}}{j_{1}!}\frac{z_{2}^{j_{2}}}{j_{2}!}\frac{%
z_{3}^{j_{3}}}{j_{3}!}\frac{z_{4}^{j_{4}}}{j_{4}!}\frac{z_{5}^{j_{5}}}{j_{5}!
}, \\
\Psi _{3}\left[\left.
\begin{array}{l}
x_{1},x_{2},x_{3} \\
x_{4}x_{5}
\end{array}\right
|z_{1};z_{2};z_{3};z_{4};z_{5}\right] &=&{\sum_{j_{1},..,j_{5}=0}^{\infty}}\frac{
(x_{1})_{j_{1}+j_{2}+j_{3}+j_{4}+j_{5}}(x_{2})_{j_{1}+j_{2}+j_{4}}(x_{3})_{j_{1}+j_{3}+j_{5}}
}{(x_{4})_{j_{1}+j_{4}+j_{5}}(x_{5})_{j_{1}+j_{2}+j_{3}}} \\
&&\times \frac{z_{1}^{j_{1}}}{j_{1}!}\frac{z_{2}^{j_{2}}}{j_{2}!}\frac{
z_{3}^{j_{3}}}{j_{3}!}\frac{z_{4}^{j_{4}}}{j_{4}!}\frac{z_{5}^{j_{5}}}{j_{5}!
}, \\
\Psi _{4}\left[\left.
\begin{array}{l}
x_{1},x_{2},x_{3} \\
x_{4},x_{5}
\end{array}\right
|z_{1};z_{2};z_{3};z_{4};z_{5}\right] &=&{\sum_{j_{1},..,j_{5}=0}^{\infty}}\frac{
(x_{1})_{j_{1}+j_{2}-j_{3}}(x_{2})_{j_{1}+j_{4}-j_{5}}(x_{3})_{-j_{2}+j_{3}+j_{4}}
}{(x_{4})_{j_{1}-j_{3}-j_{5}}(x_{5})_{-j_{2}+j_{4}-j_{5}}} \\
&&\times \frac{z_{1}^{j_{1}}}{j_{1}!}\frac{z_{2}^{j_{2}}}{j_{2}!}\frac{
z_{3}^{j_{3}}}{j_{3}!}\frac{z_{4}^{j_{4}}}{j_{4}!}\frac{z_{5}^{j_{5}}}{j_{5}!
}.
\end{eqnarray*}
The set of parameters $x_i$ and variables $z_i$ for the two and three-point
functions are listed in Appendices B and C respectively.

\section{One-loop two-point solutions}
$ $

The expressions for each of the solutions to the one-loop two-point
function, $J^{(2)}_{n}=J^{(2)}_{n}(a,b,D,m_{1},m_{2})$, where
$n=1,2,...,7,$ are given by $J^{(2)}_{n}=D_{n}F_4$, where the Appel
hypergeometric function $F_4$ is given in Appendix A and the
coefficients $D_{n}$ are shown in Table 1 below, whewreas the
parameters and variables of the $F_4$ function are listed in Table 2
following.

\vspace{.5cm}

\begin{tabular}{|c|c|}
\hline
$n$ & $D_{n}$ \\ \hline\hline
$1$  & \multicolumn{1}{|l|}{$\pi^{D/2}(q^2)^b(-m_1^2)^{a+D/2}(-a)_{-D/2}$} \\ \hline
$2$  & \multicolumn{1}{|l|}{$\pi^{D/2}(q^2)^a(-m_2^2)^{b+D/2}(-b)_{-D/2}$} \\ \hline
$3$  & \multicolumn{1}{|l|}{$\pi^{D/2}(q^2)^{a+b+D/2}(-a)_{-b-D/2}(-b)_{2b+D/2}(a+b+D)_{-b-D/2}$} \\ \hline\hline\hline
$4$  & \multicolumn{1}{|l|}{$\pi^{D/2}(-m_1^2)^a(-m_2^2)^{b+D/2}(-b)_{-D/2}$} \\ \hline
$5$  & \multicolumn{1}{|l|}{$\pi^{D/2}(-m_1^2)^{a+b+D/2}(-a)_{-b-D/2}(D/2)_b$} \\ \hline\hline\hline
$6$  & \multicolumn{1}{|l|}{$\pi^{D/2}(-m_1^2)^{a+D/2}(-m_2^2)^b(-a)_{-D/2}$} \\ \hline
$7$  & \multicolumn{1}{|l|}{$\pi^{D/2}(-m_2^2)^{a+b+D/2}(-b)_{-a-D/2}(D/2)_a$} \\ \hline
\multicolumn{2}{c}{Table-1}
\end{tabular}

\begin{tabular}{|c|c|l|}
\hline
$n$ & \multicolumn{1}{|c|}{$x_{1},x_{2},x_{3},x_{4}$} & \multicolumn{1}{|c|}{$z_{1},z_{2}$} \\ \hline\hline
$1$  & \multicolumn{1}{|l|}{$1-b-D/2,-b,1+a+D/2,1-b-D/2$}
& $\frac{m_{1}^{2}}{p^{2}};\frac{m_{2}^{2}}{p^{2}}$ \\  \hline 
$2$  & \multicolumn{1}{|l|}{$1-a-D/2,-a,1-a-D/2,1+b+D/2$}
& $\frac{m_{1}^{2}}{p^{2}};\frac{m_{2}^{2}}{p^{2}}$ \\  \hline
$3$  & \multicolumn{1}{|l|}{$1-a-b-D,-a-b-D/2,1-a-D/2,1-b-D/2$}
& $\frac{m_{1}^{2}}{p^{2}};\frac{m_{2}^{2}}{p^{2}}$ \\  \hline\hline\hline
$4$  & \multicolumn{1}{|l|}{$-a,D/2,D/2,1+b+D/2$}
& $\frac{p^{2}}{m_1^{2}};\frac{m_{2}^{2}}{m_1^{2}}$ \\  \hline
$5$  & \multicolumn{1}{|l|}{$-b,-a-b-D/2,D/2,1-b-D/2$}
& $\frac{p^{2}}{m_1^{2}};\frac{m_{2}^{2}}{m_1^{2}}$ \\  \hline\hline\hline
$6$  & \multicolumn{1}{|l|}{$D/2,-b,D/2,1+a+D/2$}
& $\frac{p^{2}}{m_2^{2}};\frac{m_{1}^{2}}{m_2^{2}}$ \\  \hline
$7$  & \multicolumn{1}{|l|}{$-a,-a-b-D/2,D/2,1-a-D/2$}
& $\frac{p^{2}}{m_2^{2}};\frac{m_{1}^{2}}{m_2^{2}}$ \\  \hline
\multicolumn{2}{c}{Table-2}
\end{tabular}

\section{One-loop three-point solutions}
$ $

The expressions for each of the solutions to the one-loop three-point
function, $J^{(3)}_{n}=J^{(3)}_{n}(a,b,c,D,m_{1},m_{2},m_{3})$, where
$n=1,2,...,69,$ are given by $J^{(3)}_{n}=D_{n}\Psi _{l},$ with
$l=1,2,3,4$. The five-fold series $\Psi_l$ are defined in Appendix A,
the coefficients $D_{n}$ are shown in Table-3 below, and the
parameters and variables of $\Psi_l$ are given in Table-4 below. The
relation between $n$ and $l$ is given by (consider
$\sigma=a+b+c+D/2$).

\[
\begin{tabular}{|c||c|c|c|c|}
\hline
$n$ & $1,...,27$ & $28,...,51$ & $52,...,63$ & $64,...,69$ \\ \hline\hline
$l$ & $1$ & $2$ & $3$ & $4$ \\ \hline
\end{tabular}
\]

\begin{tabular}{|c|c|}
\hline
$n$ & $D_{n}$ \\ \hline\hline
$1$  & \multicolumn{1}{|l|}{$\pi
^{D/2}(r^{2})^{-a-D/2}(p^{2})^{\sigma-c}(q^{2})^{\sigma-b}(-a)_{2a+D/2}(-b)_{-a-D/2}(-c)_{-a-D/2}
$} \\ \hline
$2$  & \multicolumn{1}{|l|}{$\pi
^{D/2}(q^{2})^{-b-D/2}(p^{2})^{\sigma-c}(r^{2})^{\sigma-a}(-b)_{2b+D/2}(-a)_{-b-D/2}(-c)_{-b-D/2}
$} \\ \hline
$3$  & \multicolumn{1}{|l|}{$\pi
^{D/2}(p^{2})^{-c-D/2}(q^{2})^{\sigma-b}(r^{2})^{\sigma-a}(-c)_{2c+D/2}(-a)_{-c-D/2}(-b)_{-c-D/2}
$} \\ \hline
$4$  & \multicolumn{1}{|l|}{$\pi
^{D/2}(p^{2})^{a-c}(q^{2})^{c}(-m_{2}^{2})^{\sigma-a}(-a)_{c}(-b)_{-c-D/2}
$} \\ \hline
$5$  & \multicolumn{1}{|l|}{$\pi
^{D/2}(q^{2})^{a-b}(p^{2})^{b}(-m_{3}^{2})^{\sigma-a}(-a)_{b}(-c)_{-b-D/2}
$} \\ \hline
$6$  & \multicolumn{1}{|l|}{$\pi
^{D/2}(p^{2})^{b-c}(r^{2})^{c}(-m_{1}^{2})^{\sigma-b}(-b)_{c}(-a)_{-c-D/2}
$} \\ \hline
$7$ & \multicolumn{1}{|l|}{$\pi
^{D/2}(r^{2})^{b-a}(p^{2})^{a}(-m_{3}^{2})^{\sigma-b}(-b)_{a}(-c)_{-a-D/2}
$} \\ \hline
$8$ & \multicolumn{1}{|l|}{$\pi
^{D/2}(q^{2})^{c-b}(r^{2})^{b}(-m_{1}^{2})^{\sigma-c}(-a)_{-b-D/2}(-c)_{b} $} \\ \hline
$9$ & \multicolumn{1}{|l|}{$\pi
^{D/2}(r^{2})^{c-a}(q^{2})^{a}(-m_{2}^{2})^{\sigma-c}(-b)_{-a-D/2}(-c)_{a}$} \\ \hline
$10$ & \multicolumn{1}{|l|}{$\pi
^{D/2}(p^{2})^{-c-D/2}(-m_{1}^{2})^{\sigma-b}(-m_{2}^{2})^{\sigma-a}(-c)_{2c+D/2}(-a)_{-c-D/2}(-b)_{-c-D/2}
$} \\ \hline
$11$ & \multicolumn{1}{|l|}{$\pi
^{D/2}(q^{2})^{-b-D/2}(-m_{1}^{2})^{\sigma-c}(-m_{3}^{2})^{\sigma-a}(-b)_{2b+D/2}(-a)_{-b-D/2}(-c)_{-b-D/2}
$} \\ \hline
$12$  & \multicolumn{1}{|l|}{$\pi
^{D/2}(r^{2})^{-a-D/2}(-m_{2}^{2})^{\sigma-c}(-m_{3}^{2})^{\sigma-b}(-a)_{2a+D/2}(-b)_{-a-D/2}(-c)_{-a-D/2}$} \\ \hline
$13$ & \multicolumn{1}{|l|}{$\pi
^{D/2}(p^{2})^{b}(-m_{1}^{2})^{a-b}(-m_{3}^{2})^{\sigma-a}(-a)_{b}(-c)_{-b-D/2}
$} \\ \hline
$14$ & \multicolumn{1}{|l|}{$\pi
^{D/2}(p^{2})^{a}(-m_{2}^{2})^{b-a}(-m_{3}^{2})^{\sigma-b}(-b)_{a}(-c)_{-a-D/2}$} \\ \hline
$15$ & \multicolumn{1}{|l|}{$\pi
^{D/2}(q^{2})^{c}(-m_{1}^{2})^{a-c}(-m_{2}^{2})^{\sigma-a}(-a)_{c}(-b)_{-c-D/2}$} \\ \hline
$16$  & \multicolumn{1}{|l|}{$\pi
^{D/2}(q^{2})^{a}(-m_{2}^{2})^{\sigma-c}(-m_{3}^{2})^{c-a}(-b)_{-a-D/2}(-c)_{a}
$} \\ \hline
$17$ & \multicolumn{1}{|l|}{$\pi
^{D/2}(r^{2})^{c}(-m_{1}^{2})^{\sigma-b}(-m_{2}^{2})^{b-c}(-a)_{-c-D/2}(-b)_{c}$} \\ \hline
$18$  & \multicolumn{1}{|l|}{$\pi
^{D/2}(r^{2})^{b}(-m_{1}^{2})^{\sigma-c}(-m_{3}^{2})^{c-b}(-a)_{-b-D/2}(-c)_{b}
$} \\ \hline
$19$ & \multicolumn{1}{|l|}{$\pi ^{D/2}(p^{2})^{\sigma-c}(-m_{3}^{2})^{c}\frac{(-b)_{2b+D/2}(-a)_{-b-D/2}}{(a+D/2)_{b+D/2}}
$} \\ \hline
$20 $  & \multicolumn{1}{|l|}{$\pi ^{D/2}(q^{2})^{\sigma-b}(-m_{2}^{2})^{b}\frac{(-c)_{2c+D/2}(-a)_{-c-D/2}}{(a+D/2)_{c+D/2}}$} \\ \hline
$21$  & \multicolumn{1}{|l|}{$\pi ^{D/2}(r^{2})^{\sigma-a}(-m_{1}^{2})^{a}\frac{(-c)_{2c+D/2}(-b)_{-c-D/2}}{(b+D/2)_{c+D/2}}
$} \\ \hline
\end{tabular}

\begin{tabular}{|c|c|}
\hline
$n$  & $D_{n}$ \\ \hline\hline
$22$ & \multicolumn{1}{|l|}{$\pi ^{D/2}(p^{2})^{b}(-m_{3}^{2})^{\sigma-b}\frac{
(-a)_{2a+D/2}(-c)_{-a-D/2}}{(b-a)_{a+D/2}} $} \\ \hline
$23$ & \multicolumn{1}{|l|}{$\pi ^{D/2}(p^{2})^{a}(-m_{3}^{2})^{\sigma-a}\frac{(-b)_{2b+D/2}(-c)_{-b-D/2}}{
(a-b)_{b+D/2}}$} \\ \hline
$24$  & \multicolumn{1}{|l|}{$\pi ^{D/2}(q^{2})^{c}(-m_{2}^{2})^{\sigma-c}\frac{(-a)_{2a+D/2}(-b)_{-a-D/2}}{(-a+c)_{a+D/2}}$} \\ \hline
$25$  & \multicolumn{1}{|l|}{$\pi ^{D/2}(q^{2})^{a}(-m_{2}^{2})^{\sigma-a}\frac{(-c)_{2c+D/2}(-b)_{-c-D/2}}{(a-c)_{c+D/2}}$} \\ \hline
$26$  & \multicolumn{1}{|l|}{$\pi ^{D/2}(r^{2})^{c}(-m_{1}^{2})^{\sigma-c}\frac{(-b)_{2b+D/2}(-a)_{-b-D/2}}{
(c-b)_{b+D/2}}$} \\ \hline
$27$  & \multicolumn{1}{|l|}{$\pi ^{D/2}(r^{2})^{b}(-m_{1}^{2})^{\sigma-b}\frac{(-c)_{2c+D/2}(-a)_{-c-D/2}}{
(b-c)_{c+D/2}} $} \\ \hline
$28 $  & \multicolumn{1}{|l|}{$\pi ^{D/2}(q^{2})^{c}(p^{2})^{\sigma-c}\frac{(-a)_{-b-D/2}(-b)_{2b+c+D/2}}{
(a+D/2)_{b+c+D/2}}  $} \\ \hline
$29$  & \multicolumn{1}{|l|}{$\pi^{D/2}(p^{2})^{b}(q^{2})^{\sigma-b}\frac{(-a)_{-c-D/2}(-c)_{b+2c+D/2}}{
(a+D/2)_{b+c+D/2}}
$} \\ \hline
$30 $  & \multicolumn{1}{|l|}{$\pi ^{D/2}(r^{2})^{c}(p^{2})^{\sigma-c}\frac{(-b)_{-a-D/2}(-a)_{2a+c+D/2}}{
(b+D/2)_{a+c+D/2}}  $} \\ \hline
$31 $  & \multicolumn{1}{|l|}{$\pi ^{D/2}(p^{2})^{a}(r^{2})^{\sigma-a}\frac{(-b)_{-c-D/2}(-c)_{a+2c+D/2}}{
(b+D/2)_{a+c+D/2}}
$} \\ \hline
$32$  & \multicolumn{1}{|l|}{$\pi ^{D/2}(q^{2})^{\sigma-b}(r^{2})^{b}\frac{(-c)_{-a-D/2}(-a)_{2a+b+D/2}}{
(c+D/2)_{a+b+D/2}}
$} \\ \hline
$33$  & \multicolumn{1}{|l|}{$\pi ^{D/2}(q^{2})^{a}(r^{2})^{\sigma-a}\frac{(-c)_{-b-D/2}(-b)_{a+2b+D/2}}{
(c+D/2)_{a+b+D/2}} $} \\ \hline
$34$  & \multicolumn{1}{|l|}{$\pi ^{D/2}(p^{2})^{b}(-m_{1}^{2})^{a+D/2}(-m_{3}^{2})^{c}(-a)_{-D/2}
$} \\ \hline
$35$  & \multicolumn{1}{|l|}{$\pi ^{D/2}(p^{2})^{a}(-m_{2}^{2})^{b+D/2}(-m_{3}^{2})^{c}(-b)_{-D/2}$} \\ \hline
$36$  & \multicolumn{1}{|l|}{$\pi ^{D/2}(q^{2})^{c}(-m_{1}^{2})^{a+D/2}(-m_{2}^{2})^{b}(-a)_{-D/2} $} \\ \hline
$37$  & \multicolumn{1}{|l|}{$\pi ^{D/2}(q^{2})^{a}(-m_{3}^{2})^{c+D/2}(-m_{2}^{2})^{b}(-c)_{-D/2}$} \\ \hline
$38$  & \multicolumn{1}{|l|}{$\pi ^{D/2}(r^{2})^{c}(-m_{2}^{2})^{b+D/2}(-m_{1}^{2})^{a}(-b)_{-D/2}$} \\ \hline
$39$  & \multicolumn{1}{|l|}{$\pi ^{D/2}(r^{2})^{b}(-m_{3}^{2})^{c+D/2}(-m_{1}^{2})^{a}(-c)_{-D/2}
$} \\ \hline
$40$  & \multicolumn{1}{|l|}{$\pi ^{D/2}(p^{2})^{b}(-m_{1}^{2})^{\sigma-b}(b+D/2)_{c}(-a)_{-c-D/2}
$} \\ \hline
$41$  & \multicolumn{1}{|l|}{$\pi ^{D/2}(p^{2})^{a}(-m_{2}^{2})^{\sigma-a}(a+D/2)_{c}(-b)_{-c-D/2}$} \\ \hline
$42$  & \multicolumn{1}{|l|}{$\pi ^{D/2}(q^{2})^{c}(-m_{1}^{2})^{\sigma-c}(c+D/2)_{b}(-a)_{-b-D/2} $} \\ \hline
$43$  & \multicolumn{1}{|l|}{$\pi ^{D/2}(q^{2})^{a}(-m_{3}^{2})^{\sigma-a}(a+D/2)_{b}(-c)_{-b-D/2}
$} \\ \hline
$44$  & \multicolumn{1}{|l|}{$\pi ^{D/2}(r^{2})^{c}(-m_{2}^{2})^{\sigma-c}(c+D/2)_{a}(-b)_{-a-D/2}$} \\ \hline
$45$  & \multicolumn{1}{|l|}{$\pi ^{D/2}(r^{2})^{b}(-m_{3}^{2})^{\sigma-b}(b+D/2)_{a}(-c)_{-a-D/2}
$} \\ \hline
\end{tabular}

\begin{tabular}{|c|c|}
\hline
$n$ & $D_{n}$ \\ \hline\hline
$46$  & \multicolumn{1}{|l|}{$\pi ^{D/2}(-m_{2}^{2})^{b}(-m_{1}^{2})^{\sigma-b}(D/2)_{c}(-a)_{-c-D/2}
$} \\ \hline
$47$  & \multicolumn{1}{|l|}{$\pi ^{D/2}(-m_{1}^{2})^{a}(-m_{2}^{2})^{\sigma-a}(D/2)_{c}(-b)_{-c-D/2}
$} \\ \hline
$48$  & \multicolumn{1}{|l|}{$\pi ^{D/2}(-m_{3}^{2})^{c}(-m_{1}^{2})^{\sigma-c}(D/2)_{b}(-a)_{-b-D/2}
$} \\ \hline
$49$  & \multicolumn{1}{|l|}{$\pi ^{D/2}(-m_{1}^{2})^{a}(-m_{3}^{2})^{\sigma-a}(D/2)_{b}(-c)_{-b-D/2}
$} \\ \hline
$50$  & \multicolumn{1}{|l|}{$\pi ^{D/2}(-m_{3}^{2})^{c}(-m_{2}^{2})^{\sigma-c}(D/2)_{a}(-b)_{-a-D/2}
$} \\ \hline
$51$  & \multicolumn{1}{|l|}{$\pi ^{D/2}(-m_{2}^{2})^{b}(-m_{3}^{2})^{\sigma-b}(D/2)_{a}(-c)_{-a-D/2}
$} \\ \hline
$52$  & \multicolumn{1}{|l|}{$\pi ^{D/2}(p^{2})^{b}(q^{2})^{c}(-m_{1}^{2})^{a+D/2}(-a)_{-D/2}
$} \\ \hline
$53$  & \multicolumn{1}{|l|}{$\pi ^{D/2}(p^{2})^{a}(r^{2})^{c}(-m_{2}^{2})^{b+D/2}(-b)_{-D/2}
$} \\ \hline
$54$  & \multicolumn{1}{|l|}{$\pi ^{D/2}(q^{2})^{a}(r^{2})^{b}(-m_{3}^{2})^{c+D/2}(-c)_{-D/2}$} \\ \hline
$55$  & \multicolumn{1}{|l|}{$\pi ^{D/2}(p^{2})^{\sigma}\frac{(-b)_{2b+c+D/2}(-a)_{-b-c-D/2}}{%
(a+c+D/2)_{b+D/2}}
$} \\ \hline
$56$  & \multicolumn{1}{|l|}{$\pi ^{D/2}(q^{2})^{\sigma}\frac{(-c)_{b+2c+D/2}(-a)_{-b-c-D/2}}{
(a+b+D/2)_{c+D/2}}
$} \\ \hline
$57$  & \multicolumn{1}{|l|}{$\pi ^{D/2}(r^{2})^{\sigma}\frac{(-c)_{a+2c+D/2}(-b)_{-a-c-D/2}}{
(a+b+D/2)_{c+D/2}}
$} \\ \hline
$58$  & \multicolumn{1}{|l|}{$\pi ^{D/2}(-m_{1}^{2})^{a+D/2}(-m_{2}^{2})^{b}(-m_{3}^{2})^{c}(-a)_{-D/2}$} \\ \hline
$59$  & \multicolumn{1}{|l|}{$\pi ^{D/2}(-m_{2}^{2})^{b+D/2}(-m_{1}^{2})^{a}(-m_{3}^{2})^{c}(-b)_{-D/2}$} \\ \hline
$60$  & \multicolumn{1}{|l|}{$\pi ^{D/2}(-m_{3}^{2})^{c+D/2}(-m_{1}^{2})^{a}(-m_{2}^{2})^{b}(-c)_{-D/2}
$} \\ \hline
$61$  & \multicolumn{1}{|l|}{$\pi ^{D/2}(-m_{1}^{2})^{\sigma}(D/2)_{b+c}(-a)_{-b-c-D/2}
$} \\ \hline
$62$  & \multicolumn{1}{|l|}{$\pi ^{D/2}(-m_{2}^{2})^{\sigma}(D/2)_{a+c}(-b)_{-a-c-D/2} $} \\ \hline
$63$  & \multicolumn{1}{|l|}{$\pi ^{D/2}(-m_{3}^{2})^{\sigma}(D/2)_{a+b}(-c)_{-a-b-D/2}$} \\ \hline
$64$  & \multicolumn{1}{|l|}{$\pi
^{D/2}(q^{2})^{\sigma-b}(p^{2})^{-c-D/2}(-m_{2}^{2})^{\sigma-a}(-a)_{-c-D/2}(-b)_{-c-D/2}(-c)_{2c+D/2}$} \\ \hline
$65$  & \multicolumn{1}{|l|}{$\pi
 ^{D/2}(p^{2})^{\sigma-c}(q^{2})^{-b-D/2}(-m_{3}^{2})^{\sigma-a}(-a)_{-b-D/2}(-c)_{-b-D/2}(-b)_{2b+D/2}$} \\ \hline
$66$  & \multicolumn{1}{|l|}{$\pi
 ^{D/2}(r^{2})^{\sigma-a}(p^{2})^{-c-D/2}(-m_{1}^{2})^{\sigma-b}(-a)_{-c-D/2}(-b)_{-c-D/2}(-c)_{2c+D/2}$} \\ \hline
$67$  & \multicolumn{1}{|l|}{$\pi
 ^{D/2}(p^{2})^{\sigma-c}(r^{2})^{-a-D/2}(-m_{3}^{2})^{\sigma-b}(-b)_{-a-D/2}(-c)_{-a-D/2}(-a)_{2a+D/2}$} \\ \hline
$68$  & \multicolumn{1}{|l|}{$\pi
^{D/2}(r^{2})^{\sigma-a}(q^{2})^{-b-D/2}(-m_{1}^{2})^{\sigma-c}(-a)_{-b-D/2}(-c)_{-b-D/2}(-b)_{2b+D/2}
$} \\ \hline
$69$  & \multicolumn{1}{|l|}{$\pi
^{D/2}(q^{2})^{\sigma-b}(r^{2})^{-a-D/2}(-m_{2}^{2})^{\sigma-c}(-b)_{-a-D/2}(-c)_{-a-D/2}(-a)_{2a+D/2}
$} \\ \hline
\multicolumn{2}{c}{Table-3}
\end{tabular}

\begin{tabular}{|c|c|l|}
\hline
$n$ & \multicolumn{1}{|c|}{$x_{1},x_{2},x_{3};\;x_{4},x_{5}$} & \multicolumn{1}{|c|}{$z_{1},z_{2},z_{3},z_{4},z_{5}$} \\
\hline\hline
$1$  & \multicolumn{1}{|l|}{$1\!-\!\sigma\!-\!D/2,-\!\sigma\!+\!c,-\!\sigma\!+\!b;\;1\!-\!a\!-\!D/2,1\!-\!\sigma\!-\!D/2$}
& $-%
\frac{m_{1}^{2}r^{2}}{p^{2}q^{2}};\frac{m_{2}^{2}}{p^{2}};\frac{m_{3}^{2}}{%
q^{2}};-\frac{p^{2}}{r^{2}};-\frac{q^{2}}{r^{2}}$ \\  \hline
$2$  & \multicolumn{1}{|l|}{$1\!-\!\sigma\!-\!D/2,-\!\sigma\!+\!c,-\!\sigma\!+\!a;\;1\!-\!b\!-\!D/2,1\!-\!\sigma\!-\!D/2$}
& $-%
\frac{m_{2}^{2}q^{2}}{p^{2}r^{2}};\frac{m_{1}^{2}}{%
p^{2}};\frac{m_{3}^{2}}{r^{2}};-\frac{p^{2}}{q^{2}};-\frac{r^{2}}{q^{2}}$ \\  \hline
$3$  & \multicolumn{1}{|l|}{$1\!-\!\sigma\!-\!D/2,-\!\sigma\!+\!b,-\!\sigma\!+\!a;\;1\!-\!c\!-\!D/2,1\!-\!\sigma\!-\!D/2$}
& $-%
\frac{m_{3}^{2}p^{2}}{q^{2}r^{2}};\frac{m_{1}^{2}}{q^{2}};\frac{m_{2}^{2}}{
r^{2}};-\frac{q^{2}}{p^{2}};-\frac{r^{2}}{p^{2}}$ \\  \hline
$4$  & \multicolumn{1}{|l|}{$-c,-\sigma+a,a+D/2;\;a+D/2,1+a-c$}
& $-\frac{p^{2}r^{2}}{m_{2}^{2}q^{2}};\frac{m_{3}^{2}p^{2}%
}{m_{2}^{2}q^{2}};\frac{p^{2}}{q^{2}};-\frac{m_{1}^{2}}{p^{2}};-\frac{m_{2}^{2}}{p^{2}}$ \\  \hline
$5$  & \multicolumn{1}{|l|}{$-b,-\sigma+a,a+D/2;\;a+D/2,1+a-b$}
& $-
\frac{q^{2}r^{2}}{m_{3}^{2}p^{2}};\frac{m_{2}^{2}q^{2}}{
m_{3}^{2}p^{2}};\frac{q^{2}}{p^{2}};-\frac{m_{1}^{2}}{q^{2}};-\frac{m_{3}^{2}}{q^{2}}$ \\  \hline
$6$  & \multicolumn{1}{|l|}{$-c,-\sigma+b,b+D/2;\;b+D/2,
1-c+b$}
& $-\frac{p^{2}q^{2}}{m_{1}^{2}r^{2}};\frac{m_{3}^{2}p^{2}}{%
m_{1}^{2}r^{2}};\frac{p^{2}}{r^{2}};-\frac{m_{1}^{2}}{p^{2}};-\frac{m_{2}^{2}%
}{p^{2}}$ \\  \hline
$7$ & \multicolumn{1}{|l|}{$-a,-\sigma+b,b+D/2;\;b+D/2,%
1-a+b$}
& $-\frac{q^{2}r^{2}}{m_{3}^{2}p^{2}};\frac{m_{1}^{2}r^{2}}{m_{3}^{2}p^{2}%
};\frac{r^{2}}{p^{2}};-\frac{m_{3}^{2}}{r^{2}};-\frac{m_{2}^{2}}{r^{2}}$ \\  \hline
$8$ & \multicolumn{1}{|l|}{$-b,-\sigma+c,c+D/2;\;c+D/2,1-b+c$}
& $-\frac{p^{2}q^{2}}{
m_{1}^{2}r^{2}};\frac{m_{2}^{2}q^{2}}{m_{1}^{2}r^{2}};\frac{q^{2}}{r^{2}};-%
\frac{m_{1}^{2}}{q^{2}};-\frac{m_{3}^{2}}{q^{2}}$ \\  \hline
$9$ & \multicolumn{1}{|l|}{$-a,-\sigma+c,c+D/2;\;c+D/2,1-a+c$}
& $-\frac{p^{2}r^{2}}{
m_{2}^{2}q^{2}};\frac{m_{1}^{2}r^{2}}{m_{2}^{2}q^{2}};\frac{r^{2}}{q^{2}};-
\frac{m_{2}^{2}}{r^{2}};-\frac{m_{3}^{2}}{r^{2}}$ \\  \hline
$10$ & \multicolumn{1}{|l|}{$-c,-\sigma+b,-\sigma+a;\;1-c-D/2,-c$}
& $-
\frac{m_{3}^{2}p^{2}}{m_{1}^{2}m_{2}^{2}};\frac{q^{2}}{m_{1}^{2}};\frac{
r^{2}}{m_{2}^{2}};-\frac{m_{1}^{2}}{p^{2}};-\frac{m_{2}^{2}}{p^{2}}$ \\  \hline
$11$ & \multicolumn{1}{|l|}{$-b,-\sigma+c,-\sigma+a;\;1-b-D/2,
-b$}
& $-\frac{m_{2}^{2}q^{2}}{m_{1}^{2}m_{3}^{2}};\frac{p^{2}}{
m_{1}^{2}};\frac{r^{2}}{m_{3}^{2}};-\frac{m_{1}^{2}}{q^{2}};-\frac{m_{3}^{2}
}{q^{2}}$ \\  \hline
$12$ & \multicolumn{1}{|l|}{$-a,-\sigma+c,-\sigma+b;\;1-a-D/2,-a$}
& $-\frac{
m_{1}^{2}r^{2}}{m_{2}^{2}m_{3}^{2}};\frac{p^{2}}{m_{2}^{2}};\frac{q^{2}}{
m_{3}^{2}};-\frac{m_{2}^{2}}{r^{2}};-\frac{m_{3}^{2}}{r^{2}}$ \\  \hline
$13$ & \multicolumn{1}{|l|}{$-b,1-b-D/2,-\sigma+a;\;1-b-D/2,1-b+a$}
& $-\frac{
m_{1}^{2}m_{2}^{2}}{m_{3}^{2}p^{2}};\frac{m_{1}^{2}}{p^{2}};\frac{
m_{1}^{2}r^{2}}{m_{3}^{2}p^{2}};-\frac{q^{2}}{m_{1}^{2}};-\frac{m_{3}^{2}}{
m_{1}^{2}}$ \\  \hline
$14$ & \multicolumn{1}{|l|}{$-a,1-a-D/2,-\sigma+b;\;1-a-D/2,1-a+b$}
& $-\frac{
m_{1}^{2}m_{2}^{2}}{m_{3}^{2}p^{2}};\frac{m_{2}^{2}}{p^{2}};\frac{%
m_{2}^{2}q^{2}}{m_{3}^{2}p^{2}};-\frac{r^{2}}{m_{2}^{2}};-\frac{m_{3}^{2}}{
m_{2}^{2}}$ \\  \hline
$15$ & \multicolumn{1}{|l|}{$-c,1-c-D/2,-\sigma+a;\;1-c-D/2,1-c+a$}
& $
-\frac{m_{1}^{2}m_{3}^{2}}{m_{2}^{2}q^{2}};\frac{m_{1}^{2}}{q^{2}};
\frac{m_{1}^{2}r^{2}}{m_{2}^{2}q^{2}};-\frac{p^{2}}{m_{1}^{2}};-\frac{
m_{2}^{2}}{m_{1}^{2}}$ \\  \hline
$16$ & \multicolumn{1}{|l|}{$-a,1-a-D/2,-\sigma+c;\;1-a-D/2,1-a+c$}
& $-\frac{
m_{1}^{2}m_{3}^{2}}{m_{2}^{2}q^{2}};
\frac{m_{3}^{2}}{q^{2}};\frac{m_{3}^{2}p^{2}}{m_{2}^{2}q^{2}};-\frac{r^{2}}{m_{3}^{2}};-\frac{m_{2}^{2}}{m_{3}^{2}}$ \\  \hline
$17$ & \multicolumn{1}{|l|}{$-c,1-c-D/2,-\sigma+b;\;1-c-D/2,1+b-c$}
& $-\frac{
m_{2}^{2}m_{3}^{2}}{m_{1}^{2}r^{2}};\frac{m_{2}^{2}}{r^{2}};\frac{
m_{2}^{2}q^{2}}{m_{1}^{2}r^{2}};-\frac{p^{2}}{m_{2}^{2}};-\frac{m_{1}^{2}}{m_{2}^{2}}$ \\  \hline
$18$ & \multicolumn{1}{|l|}{$-b,1-b-D/2,-\sigma+c;\;1-b-D/2,1-b+c$}
& $
-\frac{m_{2}^{2}m_{3}^{2}}{m_{1}^{2}r^{2}};\frac{m_{3}^{2}}{r^{2}};\frac{m_{3}^{2}p^{2}}{
m_{1}^{2}r^{2}};-\frac{
q^{2}}{m_{3}^{2}};-\frac{m_{1}^{2}}{m_{3}^{2}}$ \\  \hline
$19$ & \multicolumn{1}{|l|}{$-c,b+D/2,a+D/2;\;1+\sigma-c,a+b+D$}
& $
-\frac{p^{2}}{m_{3}^{2}};\frac{q^{2}}{m_{3}^{2}};\frac{r^{2}}{m_{3}^{2}};-
\frac{m_{2}^{2}}{p^{2}};-\frac{m_{1}^{2}}{p^{2}}$ \\  \hline
$20$ & \multicolumn{1}{|l|}{$-b,c+D/2,a+D/2;\;1+\sigma-b,
a+c+D$}
& $-\frac{q^{2}}{m_{2}^{2}};\frac{p^{2}}{m_{2}^{2}};\frac{r^{2}}{
m_{2}^{2}};-\frac{m_{3}^{2}}{q^{2}};-\frac{m_{1}^{2}}{q^{2}}$ \\  \hline
$21$ & \multicolumn{1}{|l|}{$-a,c+D/2,b+D/2;\;1+\sigma-a,
b+c+D$}
& $-\frac{r^{2}}{m_{1}^{2}};\frac{p^{2}}{m_{1}^{2}};\frac{q^{2}}{
m_{1}^{2}};-\frac{m_{3}^{2}}{r^{2}};-\frac{m_{2}^{2}}{r^{2}}$ \\  \hline
\end{tabular}

\begin{tabular}{|c|c|l|}
\hline
$n$ & \multicolumn{1}{|c|}{$x_{1},x_{2},x_{3},x_{4},x_{5}$} & \multicolumn{1}{|c|}{$z_{1},z_{2},z_{3},z_{4},z_{5}$} \\
\hline\hline
$22$ & \multicolumn{1}{|l|}{$-b,1-b-D/2,a+D/2;\;1+\sigma-b,1-b+a$}
& $-
\frac{m_{3}^{2}}{p^{2}};\frac{m_{2}^{2}}{p^{2}};\frac{r^{2}}{p^{2}};-\frac{
q^{2}}{m_{3}^{2}};-\frac{m_{1}^{2}}{m_{3}^{2}}$ \\  \hline
$23$ & \multicolumn{1}{|l|}{$-a,1-a-D/2,b+D/2;\;1+\sigma-a,1-a+b$}
& $-\frac{
m_{3}^{2}}{p^{2}};\frac{m_{1}^{2}}{p^{2}};\frac{q^{2}}{p^{2}};-\frac{r^{2}}{
m_{3}^{2}};-\frac{m_{2}^{2}}{m_{3}^{2}}$ \\  \hline
$24$ & \multicolumn{1}{|l|}{$-c,1-c-D/2,a+D/2;\;1+\sigma-c,1-c+a$}
& $-\frac{
m_{2}^{2}}{q^{2}};\frac{m_{3}^{2}}{q^{2}};\frac{r^{2}}{q^{2}};-\frac{p^{2}}{
m_{2}^{2}};-\frac{m_{1}^{2}}{m_{2}^{2}}$ \\  \hline
$25$ & \multicolumn{1}{|l|}{$-a,1-a-D/2,c+D/2;\;1+\sigma-a,1-a+c$}
& $-
\frac{m_{2}^{2}}{q^{2}};\frac{m_{1}^{2}}{q^{2}};\frac{p^{2}}{q^{2}};-\frac{r^{2}}{m_{2}^{2}};-\frac{
m_{3}^{2}}{m_{2}^{2}}$ \\  \hline
$26$ & \multicolumn{1}{|l|}{$-c,1-c-D/2,b+D/2;\;1+\sigma-c,1+b-c$}
& $-
\frac{m_{1}^{2}}{r^{2}};\frac{m_{3}^{2}}{r^{2}};\frac{q^{2}}{r^{2}};-\frac{
p^{2}}{m_{1}^{2}};-\frac{m_{2}^{2}}{m_{1}^{2}}$ \\  \hline
$27$ & \multicolumn{1}{|l|}{$-b,1-b-D/2,c+D/2;\;
1+\sigma-b,1-b+c$}
& $-\frac{m_{1}^{2}}{r^{2}};\frac{
m_{2}^{2}}{r^{2}};\frac{p^{2}}{r^{2}};-\frac{q^{2}}{m_{1}^{2}};-\frac{m_{3}^{2}}{m_{1}^{2}}$ \\  \hline
$28$ & \multicolumn{1}{|l|}{$1-\sigma-D/2,-c,-\sigma+c;\;1-a-D/2,1-\sigma+a$}
& $\frac{m_{2}^{2}}{p^{2}};\frac{m_{1}^{2}}{p^{2}};-\frac{r^{2}}{q^{2}};\frac{
p^{2}}{q^{2}};\frac{m_{3}^{2}}{q^{2}}$ \\  \hline
$29$ & \multicolumn{1}{|l|}{$1-\sigma-D/2,-b,-\sigma+b;\;1-a-D/2,1-\sigma+a$}
& $\frac{m_{3}^{2}}{q^{2}};\frac{m_{1}^{2}}{q^{2}};-\frac{r^{2}}{p^{2}};\frac{
q^{2}}{p^{2}};\frac{m_{2}^{2}}{p^{2}}$ \\  \hline
$30$ & \multicolumn{1}{|l|}{$1-\sigma-D/2,-c,-\sigma+c;\;1-b-D/2,1-\sigma+b$}
& $\frac{m_{1}^{2}}{p^{2}};\frac{m_{2}^{2}}{p^{2}};-\frac{q^{2}}{r^{2}};\frac{
p^{2}}{r^{2}};\frac{m_{3}^{2}}{r^{2}}$ \\  \hline
$31$ & \multicolumn{1}{|l|}{$1-\sigma-D/2,-a,-\sigma+a;\;1-b-D/2,1-\sigma+b$}
& $\frac{
m_{3}^{2}}{r^{2}};\frac{m_{2}^{2}}{r^{2}};-\frac{q^{2}}{p^{2}};\frac{r^{2}}{
p^{2}};\frac{m_{1}^{2}}{p^{2}}$ \\  \hline
$32$ & \multicolumn{1}{|l|}{$1-\sigma-D/2,-b,-\sigma+b;\;1-c-D/2,1-\sigma+c$}
& $\frac{
m_{1}^{2}}{q^{2}};\frac{m_{3}^{2}}{q^{2}};-\frac{p^{2}}{r^{2}};\frac{q^{2}}{%
r^{2}};\frac{m_{2}^{2}}{r^{2}}$ \\  \hline
$33$ & \multicolumn{1}{|l|}{$1-\sigma-D/2,-a,-\sigma+a;\;1-c-D/2,1-\sigma+c$}
& $\frac{m_{2}^{2}}{
r^{2}};\frac{m_{3}^{2}}{r^{2}};-\frac{p^{2}}{q^{2}};\frac{r^{2}}{q^{2}};
\frac{m_{1}^{2}}{q^{2}}$ \\  \hline
$34$ & \multicolumn{1}{|l|}{$-c,-b,b+D/2;\;b+D/2,1+a+D/2$}
& $\frac{
m_{1}^{2}}{m_{3}^{2}};\frac{q^{2}}{m_{3}^{2}};-\frac{m_{1}^{2}}{p^{2}};
\frac{m_{2}^{2}}{p^{2}},\frac{m_{1}^{2}r^{2}}{m_{3}^{2}p^{2}}$ \\  \hline
$35$ & \multicolumn{1}{|l|}{$-c,-a,a+D/2;\;a+D/2,1+b+D/2$}
& $\frac{m_{2}^{2}
}{m_{3}^{2}};\frac{r^{2}}{m_{3}^{2}};-\frac{m_{2}^{2}}{p^{2}};\frac{
m_{1}^{2}}{p^{2}};\frac{m_{2}^{2}q^{2}}{m_{3}^{2}p^{2}}$ \\  \hline
$36$ & \multicolumn{1}{|l|}{$-b,-c,c+D/2;\;c+D/2,1+a+D/2$}
& $\frac{
m_{1}^{2}}{m_{2}^{2}};\frac{p^{2}}{m_{2}^{2}};-\frac{m_{1}^{2}}{q^{2}};
\frac{m_{3}^{2}}{q^{2}};\frac{m_{1}^{2}r^{2}}{m_{2}^{2}q^{2}}$ \\  \hline
$37$ & \multicolumn{1}{|l|}{$-b,-a,a+D/2;\;a+D/2,1+c+D/2$}
& $\frac{%
m_{3}^{2}}{m_{2}^{2}};\frac{r^{2}}{m_{2}^{2}};-\frac{m_{3}^{2}}{q^{2}};
\frac{m_{1}^{2}}{q^{2}};\frac{m_{3}^{2}p^{2}}{m_{2}^{2}q^{2}}$ \\  \hline
$38$ & \multicolumn{1}{|l|}{$-a,-c,c+D/2;\;c+D/2,1+b+D/2$}
& $\frac{
m_{2}^{2}}{m_{1}^{2}};\frac{p^{2}}{m_{1}^{2}};-\frac{m_{2}^{2}}{r^{2}};
\frac{m_{3}^{2}}{r^{2}};\frac{m_{2}^{2}q^{2}}{m_{1}^{2}r^{2}}$ \\  \hline
$39$ & \multicolumn{1}{|l|}{$-a,-b,b+D/2;\;b+D/2,1+c+D/2$}
& $\frac{m_{3}^{2}
}{m_{1}^{2}};\frac{q^{2}}{m_{1}^{2}};-\frac{m_{3}^{2}}{r^{2}};\frac{
m_{2}^{2}}{r^{2}};\frac{m_{3}^{2}p^{2}}{m_{1}^{2}r^{2}}$ \\  \hline
$40$ & \multicolumn{1}{|l|}{$-c,-b,-\sigma+b;\;b+D/2,1-\sigma+a$}
& $\frac{
m_{3}^{2}}{m_{1}^{2}};\frac{q^{2}}{m_{1}^{2}};-\frac{m_{2}^{2}}{p^{2}};
\frac{m_{1}^{2}}{p^{2}};\frac{r^{2}}{p^{2}}$ \\  \hline
$41$ & \multicolumn{1}{|l|}{$-c,-a,-\sigma+a;\;a+D/2,1-\sigma+b$}
& $\frac{m_{3}^{2}}{m_{2}^{2}};\frac{r^{2}}{m_{2}^2};-\frac{
m_{1}^{2}}{p^{2}};\frac{m_{2}^{2}}{p^{2}};\frac{q^{2}}{p^{2}}$ \\  \hline
$42$ & \multicolumn{1}{|l|}{$-b,-c,-\sigma+c;\;c+D/2,1-\sigma+a$}
& $\frac{
m_{2}^{2}}{m_{1}^{2}};\frac{p^{2}}{m_{1}^{2}};-\frac{m_{3}^{2}}{q^{2}};
\frac{m_{1}^{2}}{q^{2}};\frac{r^{2}}{q^{2}}$ \\  \hline
$43$ & \multicolumn{1}{|l|}{$-b,-a,-\sigma+a;\;a+D/2,1-\sigma+c$}
& $\frac{
m_{2}^{2}}{m_{3}^{2}};\frac{r^{2}}{m_{3}^{2}};-\frac{m_{1}^{2}}{q^{2}};
\frac{m_{3}^{2}}{q^{2}};\frac{p^{2}}{q^{2}}$ \\  \hline
$44$ & \multicolumn{1}{|l|}{$-a,-c,-\sigma+c;\;c+D/2,1-\sigma+b$}
& $\frac{
m_{1}^{2}}{m_{2}^{2}};\frac{p^{2}}{m_{2}^{2}};-\frac{m_{3}^{2}}{r^{2}};
\frac{m_{2}^{2}}{r^{2}};\frac{q^{2}}{r^{2}}$ \\  \hline
$45$ & \multicolumn{1}{|l|}{$-a,-b,-\sigma+b;\;b+D/2,1-\sigma+c$}
& $\frac{
m_{1}^{2}}{m_{3}^{2}};\frac{q^{2}}{m_{3}^{2}};-\frac{m_{2}^{2}}{r^{2}};
\frac{m_{3}^{2}}{r^{2}};\frac{p^{2}}{r^{2}}$ \\  \hline
\end{tabular}

\begin{tabular}{|c|c|l|}
\hline
$n$ & \multicolumn{1}{|c|}{$x_{1},x_{2},x_{3},x_{4},x_{5}$} & \multicolumn{1}{|c|}{$z_{1},z_{2},z_{3},z_{4},z_{5}$} \\
\hline\hline
$46$ & \multicolumn{1}{|l|}{$-b,-c,c+D/2;\;1+\sigma-b,D/2$}
& $\frac{
p^{2}}{m_{2}^{2}};\frac{m_{1}^{2}}{m_{2}^{2}};-\frac{q^{2}}{m_{1}^{2}};
\frac{m_{3}^{2}}{m_{1}^{2}};\frac{r^{2}}{m_{2}^{2}}$ \\  \hline
$47$ & \multicolumn{1}{|l|}{$-a,-c,c+D/2;\;1+\sigma-a,D/2$}
& $\frac{
p_{2}^{2}}{m_{1}^{2}};\frac{m_{2}^{2}}{m_{1}^{2}};-\frac{r^{2}}{m_{2}^{2}};\frac{m_{3}^{2}}{
m_{2}^{2}};\frac{q^{2}}{m_{1}^{2}}$ \\  \hline
$48$ & \multicolumn{1}{|l|}{$-c,-b,b+D/2;\;1+\sigma-c,D/2$}
& $\frac{q^{2}}{
m_{3}^{2}};\frac{m_{1}^{2}}{m_{3}^{2}};-\frac{p^{2}}{m_{1}^{2}};\frac{
m_{2}^{2}}{m_{1}^{2}},\frac{r^{2}}{m_{3}^{2}}$ \\  \hline
$49$ & \multicolumn{1}{|l|}{$-a,-b,b+D/2;\;1+\sigma-a,D/2$}
& $\frac{q^{2}}{
m_{1}^{2}};\frac{m_{3}^{2}}{m_{1}^{2}};-\frac{r^{2}}{m_{3}^{2}};\frac{
m_{2}^{2}}{m_{3}^{2}};\frac{p^{2}}{m_{1}^{2}}$ \\  \hline
$50$ & \multicolumn{1}{|l|}{$-c,-a,a+D/2;\;1+\sigma-c,D/2$}
& $\frac{r^{2}
}{m_{3}^{2}};\frac{m_{2}^{2}}{m_{3}^{2}};-\frac{p^{2}}{m_{2}^{2}};\frac{
m_{1}^{2}}{m_{2}^{2}};\frac{q^{2}}{m_{3}^{2}}$ \\  \hline
$51$ & \multicolumn{1}{|l|}{$-b,-a,a+D/2;\;1+\sigma-b,D/2$}
& $\frac{r^{2}}{
m_{2}^{2}};\frac{m_{3}^{2}}{m_{2}^{2}};-\frac{q^{2}}{m_{3}^{2}};\frac{
m_{1}^{2}}{m_{3}^{2}};\frac{p^{2}}{m_{2}^{2}}$ \\  \hline
$52$ & \multicolumn{1}{|l|}{$1-\sigma+a,-b,-c;\;1+a+D/2,1-\sigma+a$}
& $-\frac{
m_{1}^{2}r^{2}}{p^{2}q^{2}};\frac{m_{2}^{2}}{p^{2}};\frac{m_{3}^{2}}{q^{2}};
\frac{m_{1}^{2}}{p^{2}};\frac{m_{1}^{2}}{q^{2}}$ \\  \hline
$53$ & \multicolumn{1}{|l|}{$1-\sigma+b,-a,-c;\;1+b+D/2,1-\sigma+b$}
& $-\frac{
m_{2}^{2}q^{2}}{p^{2}r^{2}};\frac{m_{1}^{2}}{p^{2}};\frac{m_{3}^{2}}{r^{2}};
\frac{m_{2}^{2}}{p^{2}};\frac{m_{2}^{2}}{r^{2}}$ \\  \hline
$54$ & \multicolumn{1}{|l|}{$1-\sigma+c,-a,-b;\;1+c+D/2,1-\sigma+c$}
& $-\frac{
m_{3}^{2}p^{2}}{q^{2}r^{2}};\frac{m_{1}^{2}}{q^{2}};\frac{m_{2}^{2}}{r^{2}};
\frac{m_{3}^{2}}{q^{2}};\frac{m_{3}^{2}}{r^{2}}$ \\  \hline
$55$ & \multicolumn{1}{|l|}{$-\sigma,1-\sigma-D/2,-c;\;1-\sigma+a,1-\sigma+b$}
& $-\frac{m_{3}^{2}}{p^{2}};\frac{m_{1}^{2}}{p^{2}};\frac{q^{2}}{p^{2}};\frac{
m_{2}^{2}}{p^{2}};\frac{r^{2}}{p^{2}}$ \\  \hline

$56$ & \multicolumn{1}{|l|}{$-\sigma,1-\sigma-D/2,-b;\;1-\sigma+a,1-\sigma+c$}
& $
-\frac{m_{2}^{2}}{q^{2}};\frac{m_{1}^{2}}{q^{2}};\frac{p^{2}}{q^{2}};\frac{m_{3}^{2}}{q^{2}};\frac{
r^{2}}{q^{2}}$ \\  \hline
$57$ & \multicolumn{1}{|l|}{$-\sigma,1-\sigma-D/2,-a;\;1-\sigma+b,1-\sigma+c$}
& $-\frac{
m_{1}^{2}}{r^{2}};\frac{m_{2}^{2}}{r^{2}};\frac{p^{2}}{r^{2}};\frac{m_{3}^{2}
}{r^{2}};\frac{q^{2}}{r^{2}}$ \\  \hline
$58$ & \multicolumn{1}{|l|}{$D/2,-b,-c;\;1+a+D/2,D/2$}
& $-\frac{
m_{1}^{2}r^{2}}{m_{2}^{2}m_{3}^{2}};\frac{p^{2}}{m_{2}^{2}};\frac{q^{2}}{
m_{3}^{2}};\frac{m_{1}^{2}}{m_{2}^{2}};\frac{m_{1}^{2}}{m_{3}^{2}}$ \\  \hline
$59$ & \multicolumn{1}{|l|}{$D/2,-a,-c;\;1+b+D/2,D/2$}
& $-\frac{m_{2}^{2}q^{2}
}{m_{1}^{2}m_{3}^{2}};\frac{p^{2}}{m_{1}^{2}};\frac{r^{2}}{m_{3}^{2}}
;\frac{m_{2}^{2}}{m_{1}^{2}};\frac{m_{2}^{2}}{m_{3}^{2}}$ \\  \hline
$60$ & \multicolumn{1}{|l|}{$D/2,-a,-b;\;1+c+D/2,D/2$}
& $-\frac{m_{3}^{2}p^{2}}{
m_{1}^{2}m_{2}^{2}};\frac{q^{2}}{m_{1}^{2}};\frac{r^{2}}{m_{2}^{2}};
\frac{m_{3}^{2}}{m_{1}^{2}};\frac{m_{3}^{2}}{m_{2}^{2}}$ \\  \hline
$61$ & \multicolumn{1}{|l|}{$-\sigma,-b,-c;\;1-\sigma+a,D/2$}
& $-\frac{r^{2}
}{m_{1}^{2}};\frac{p^{2}}{m_{1}^{2}};\frac{q^{2}}{m_{1}^{2}};\frac{m_{2}^{2}%
}{m_{1}^{2}};\frac{m_{3}^{2}}{m_{1}^{2}}$ \\  \hline
$62$ & \multicolumn{1}{|l|}{$-\sigma,-a,-c;\;1-\sigma+b,D/2$}
& $-\frac{q^{2}
}{m_{2}^{2}};\frac{p^{2}}{m_{2}^{2}};\frac{r^{2}}{m_{2}^{2}};\frac{m_{1}^{2}%
}{m_{2}^{2}};\frac{m_{3}^{2}}{m_{2}^{2}}$ \\  \hline
$63$ & \multicolumn{1}{|l|}{$-\sigma,-a,-b;\;1-\sigma+c,D/2$}
& $-\frac{p^{2}
}{m_{3}^{2}};\frac{q^{2}}{m_{3}^{2}};\frac{r^{2}}{m_{3}^{2}};\frac{m_{1}^{2}%
}{m_{3}^{2}};\frac{m_{2}^{2}}{m_{3}^{2}}$ \\  \hline
$64$ & \multicolumn{1}{|l|}{$-\sigma+a,a+D/2,c+D/2;\;a+D/2,1+\sigma-b$}
& $
\frac{r^{2}}{m_{2}^{2}};\frac{m_{3}^{2}p^{2}}{m_{2}^{2}q^{2}};\frac{m_{2}^{2}}{p^{2}}%
;\frac{q^{2}}{p^{2}};-\frac{m_{1}^{2}}{q^{2}}$ \\  \hline
$65$ & \multicolumn{1}{|l|}{$-\sigma+a,a+D/2,b+D/2;\;a+D/2,1+\sigma-c$}
& $\frac{r^{2}}{m_{3}^{2}};\frac{m_{2}^{2}q^{2}}{m_{3}^{2}p^{2}};\frac{m_{3}^{2}}{q^{2}}
;\frac{p^{2}}{q^{2}};-\frac{m_{1}^{2}}{p^{2}}$ \\  \hline
$66$ & \multicolumn{1}{|l|}{$-\sigma+b,b+D/2,
c+D/2;\;b+D/2,1+\sigma-a$}
& $\frac{q^{2}}{m_{1}^{2}};\frac{m_{3}^{2}p^{2}}{
m_{1}^{2}r^{2}};\frac{m_{1}^{2}}{p^{2}};\frac{r^{2}}{p^{2}};-\frac{m_{2}^{2}%
}{r^{2}}$ \\  \hline
$67$ & \multicolumn{1}{|l|}{$-\sigma+b,b+D/2,a+D/2;\;b+D/2,1+\sigma-c$}
& $\frac{q^{2}}{m_{3}^{2}};\frac{m_{1}^{2}r^{2}}{m_{3}^{2}p^{2}};
\frac{m_{3}^{2}}{r^{2}};\frac{p^{2}}{r^{2}};-\frac{m_{2}^{2}}{p^{2}}$ \\  \hline
$68$ & \multicolumn{1}{|l|}{$-\sigma+c,c+D/2,
b+D/2;\;c+D/2,1+\sigma-a$}
& $\frac{p^{2}}{m_{1}^{2}};\frac{m_{2}^{2}q^{2}}{m_{1}^{2}r^{2}
};\frac{m_{1}^{2}}{q^{2}};\frac{r^{2}}{q^{2}};-\frac{m_{3}^{2}}{r^{2}}$ \\  \hline
$69$ & \multicolumn{1}{|l|}{$-\sigma+c,c+D/2,a+D/2;\;c+D/2,1+\sigma-b$}
& $\frac{p^{2}}{m_{2}^{2}};\frac{m_{1}^{2}r^{2}}{m_{2}^{2}q^{2}};\frac{
m_{2}^{2}}{r^{2}};\frac{q^{2}}{r^{2}};-\frac{m_{3}^{2}}{q^{2}}$ \\  \hline
\multicolumn{2}{c}{Table-4}
\end{tabular}

\end{document}